# Quantifying the Immediate Effects of the COVID-19 Pandemic on Scientists

Kyle R. Myers[1,2*], Wei Yang Tham[1,2], Yian Yin[3,4,5], Nina Cohodes[2,6], Jerry G. Thursby[2,7], Marie C. Thursby[2,8], Peter E. Schiffer[9], Joseph T. Walsh[5,10], Karim R. Lakhani[1,2,6], Dashun Wang[3,4,5,11*]

[1]Harvard Business School, Harvard University, Boston, MA, USA
[2]Laboratory for Innovation Science at Harvard, Harvard University, Boston, MA, USA
[3]Center for Science of Science and Innovation, Northwestern University, Evanston, IL, USA
[4]Northwestern Institute on Complex Systems, Northwestern University, Evanston, IL, USA
[5]McCormick School of Engineering, Northwestern University, Evanston, IL, USA
[6]Institute for Quantitative Social Science, Harvard University, Boston, MA, USA
[7]TyGeron Institute, Nashville, TN, USA
[8]Scheller College of Business, Georgia Institute of Technology, Atlanta, GA, USA
[9]Department of Applied Physics and Department of Physics, Yale University, New Haven, CT, USA
[10]University of Illinois System, Urbana, IL, USA
[11]Kellogg School of Management, Northwestern University, Evanston, IL, USA
[*]Correspondence to: kmyers@hbs.edu, dashun.wang@northwestern.edu

**The COVID-19 pandemic has undoubtedly disrupted the scientific enterprise, but we lack empirical evidence on the nature and magnitude of these disruptions. Here we report the results of a survey of approximately 4,500 Principal Investigators (PIs) at U.S.- and Europe-based research institutions. Distributed in mid-April 2020, the survey solicited information about how scientists' work changed from the onset of the pandemic, how their research output might be affected in the near future, and a wide range of individuals' characteristics. Scientists report a sharp decline in time spent on research on average, but there is substantial heterogeneity with a significant share reporting no change or even increases. Some of this heterogeneity is due to field-specific differences, with laboratory-based fields being the most negatively affected, and some is due to gender, with female scientists reporting larger declines. However, among the individuals' characteristics examined, the largest disruptions are connected to a usually unobserved dimension: childcare. Reporting a young dependent is associated with declines similar in magnitude to those reported by the laboratory-based fields and can account for a significant fraction of gender differences. Amidst scarce evidence about the role of parenting in scientists' work, these results highlight the fundamental and heterogeneous ways this pandemic is affecting the scientific workforce, and may have broad relevance for shaping responses to the pandemic's effect on science and beyond.**

**A large-scale survey of scientists amidst the COVID-19 pandemic**

By mid-April 2020, the cumulative number of deaths due to COVID-19 had reached approximately 115,000 with nearly 1,800 deaths per day in the U.S. and 3,000 deaths per day in Europe[1]. Throughout the U.S. and Europe, schools and workplaces were typically required to be closed and restrictions on gatherings of more than 10 people were in place in most countries[2]. For scientists, not only did this drastically change their daily lives, it severely limited the possibilities of using traditional workspaces as most institutions had suspended "non-essential" activities on campus[3–10].

To collect timely data on how the pandemic affected scientists' work, we disseminated a survey to U.S.- and Europe-based scientists across a wide range of institutions, career stages, and demographic backgrounds. We identified the corresponding authors for all journal articles indexed by the Web of Science in the past decade, and then randomly sampled 400,000 U.S.- and Europe-based email addresses (see SI S1 for more). We distributed the survey on Monday April 13th, 2020, about 1 month after the World Health Organization declared the COVID-19 pandemic. Within one week, the survey received full responses from 4,535 individuals who self-identified as faculty or PIs from academic or non-profit research institutions. Respondents were located in all 50 states in the U.S. (63.7% of the sample, Figure S1A), 35 countries in Europe (36.3% of the sample, Figure S1B), and were affiliated with the full spectrum of research fields listed in the survey. For more on the response rate, sampling method, and a comparison to a national survey of doctorate-level researchers, see SI S3.

Motivated by prior research on scientific productivity[11–15], the survey solicited information about scientists' working hours, how this time is allocated across different tasks, and how these time allocations have changed since the onset of the pandemic. We asked scientists to estimate changes to their research output—the quantity and impact of their publications—in coming years relative to prior years. We also solicited a wide range of characteristics including field of study, career stage (e.g., tenure status), demographics (e.g., age, gender, number and age of dependents in the household), and other features (e.g., institution closure and whether the respondent was exempt from any closures). Details on the survey instrument are included in SI S2, and Table S1 reports summary statistics for all the respondents used in the analyses.

**Heterogeneous effects of the pandemic**

To understand the immediate impacts of the pandemic, we compare the reported level and allocation of work hours pre-pandemic and at the time of the survey. Figures 1A and 1B illustrate two primary findings. First, there is a sharp decline in total work hours, with the average dropping from 61.4 hours per week pre-pandemic to 54.4 at the time of the survey (diff.=-6.9, s.e.=0.20). In particular, 5.01% of scientists reported that they worked 42 hours or less before the pandemic, but this share increased nearly six-fold to 29.7% by the time of the survey (diff.=24.7, s.e.=0.67). Second, there is large heterogeneity in changes across respondents. Although 55.0% reported a decline in total work hours, 27.3% reported no change, and 17.7% reported an increase in time devoted to work. This significant fraction of scientists reporting no change or increases in their work hours is notable given that 91.0% of respondents reported their institution was closed for non-essential personnel.

To decompose these changes, we compare scientists' reported time allocations across four broad categories of work: research (e.g., planning experiments, collecting or analyzing data, writing), fundraising (e.g., writing grant proposals), teaching, and all other tasks (e.g., administrative, editorial, or clinical duties). We find that among the four categories, research activities have seen the largest negative changes. Whereas total work hours decrease by 11.3% on average, research hours have declined by 24.4% (teaching, fundraising, and "all other tasks" decrease by 1.9%, 9.3%, and 0.7%, respectively). Comparing the share of time allocated across the tasks (Figure 1C-F), we find that research is the only category that sees an overall decline in the share of time committed (median changes: -16.2% for research, 0% for fundraising, +2.7% for teaching, and +2.0% for all other tasks). Overall, these results indicate that scientists' research time has been disrupted the most, and the declines in time spent on the other three categories are mainly due to the decline in total work hours. Furthermore, correlations suggest that research may be a substitute for each of the three other tasks (see SI S5.1 and Figure S4). Still, despite the large negative changes in research time, substantial heterogeneity remains, as 9.4% reported no change and 21.2% reported spending more time on research. The sizable heterogeneity begs the question as to what factors are most responsible for the observed heterogeneous effects among scientists.

**Unpacking field- and individual-level effects**

To unpack the varied effects of the pandemic, we first examine across-field differences. Figure 2A depicts the average change in reported research time across the 20 different fields we surveyed. Fields that tend to rely on physical laboratories and time-sensitive experiments—such as biochemistry, biological sciences, chemistry and chemical engineering—report the largest declines in research time, in the range of 30-40% below pre-pandemic levels. Conversely, fields that are less equipment-intensive—such as mathematics, statistics, computer science, and economics—report the lowest average declines in research time. The difference between fields can be as large as four-fold, again highlighting the heterogeneity in how certain scientists are being affected.

These field-level differences may be due to the nature of work specific to each field, but may also be due to differences in the characteristics of individuals that work in each field. To untangle these factors, we use a Lasso regression approach to select amongst (1) a vector of field indicator variables, and (2) a vector of flexible transformations of demographic controls and pre-pandemic features (e.g., research funding level, time allocations before the pandemic). The Lasso is a data-driven approach to feature selection that minimizes overfitting by selecting only variables with significant explanatory power[16,17]. We then regress the reported change in research time on the Lasso-selected variables in a post-Lasso regression, allowing us to estimate conditional associations for each variable selected (see SI S4).

Comparing Figure 2A and 2B, we find that the contrast between the "laboratory" or "bench science" fields versus the more computational or theoretical fields is still significant in the post-Lasso regression, indicating that differences inherent to these fields are likely important mediators of how the pandemic is affecting scientists. Although we cannot reject a null hypothesis of no change, there is also suggestive evidence of an increase in research time for the health sciences, possibly due to work related to COVID-19. Importantly, we also find that most of the variation across fields is diminished once we condition on the individual-level features selected by the Lasso, which suggests a large amount of heterogeneity is due to these individual-level differences. Indeed, the standard deviation of the twenty field-level averages of reported changes in research time is 7.4%. By contrast, the standard deviation of the individual-level residuals from these field-level averages—that is, how much each individual's response differs from the average in their

field—is 50.5%, indicating there is substantial variation across individuals even within the same field.

To illustrate the raw individual-level variation, we measure the average change in reported research time across demographic and other group features (Figure 2C). Given the persistent gender gap in science[18–28], we include interactions with the female indicator to explore potential gender-specific differences. We find that there are indeed widespread changes across the range of individual-level features we examined. Yet, when we use the Lasso and regression to control for the field differences documented in Figure 2A, we find marked changes in the relevance of certain individual-level features.

Figure 2D plots the post-Lasso regression coefficients associated with the demographic and career-stage characteristics and reveals four main results. First, career stage appears to be a poor predictor of the impacts of the pandemic, as conditional changes in research time for older versus younger and tenured versus untenured faculty are statistically indistinguishable. Second, scientists who report being subject to a facility closure also report only minor unconditional differences in their research time (Figure 2C), and this feature is not selected by the Lasso as a relevant predictor for changes in research time. Third, there is a clear gender difference. Holding field and all other observable features fixed, female scientists report a 4.2% larger decline in research time (s.e.=1.5). Fourth, child dependent care is associated with the largest effect. Reporting a dependent under 5 years old is associated with a 15.8% (s.e.=2.1) larger decline in research time, showing a substantially larger effect than any other individual-level features.

Reporting a dependent 6 to 11 years old is also associated with a negative impact, *ceteris paribus*, but that decline is smaller than the decline associated with dependents under 5 years old. This is consistent with shifts in the demands of childcare as children age. Having multiple dependents is associated with an additional 4.5% decline (s.e.=1.6) in research time. Overall, these results are consistent with preliminary reports of differential declines in female scientists' productivity during the pandemic[29,30]. Our findings further indicate that some of the gender discrepancy can be attributed to female scientists being more likely to have young children as dependents (21.2% of female scientists in our sample report having dependents under the age of 5, compared to 17.7% of male and other scientists, s.e. of diff.=3.6). For further results related to the other three task categories, see SI S5.2.

**Changes in forecasted publication output**

To estimate the potential downstream impact of the pandemic, we also asked respondents to forecast how their research publication output in 2020 and 2021—in terms of the quantity and impact of their publications—will compare to their output in 2018 and 2019. We randomly assigned respondents to make a forecast for one of six possible scenarios where they were to take as given the duration of the pandemic to be 1, 2, 3, 4, 6, or 8 months from the time of the survey. For more on how we use this introduced random variation and adjust scientists' forecasts to account for underlying trends in publication output, see SI S4.2.

Figure 3A plots the distribution of the estimated changes in publication quantity and impact due to the pandemic. We find that, on average, quantity is projected to decline 13.0% (s.d.=37.7). For comparison, prior estimates show that in the biomedical sciences, receiving a grant of approximately one million dollars from the National Institutes of Health raises a PI's short run publication output by 7-12%[31,32], suggesting that a projected decline of 13% is not negligible. Moreover, the decline in output is not limited to quantity, as impact is projected to decline by 7.9% on average (s.d.=31.0).

To understand which scientists are most likely to forecast larger declines in their output due to the pandemic, we repeat the Lasso-based regression approach using these forecasts as dependent variables. These analyses uncover two notable findings (Figure 3B). First, all of the features selected as relevant are related to caring for dependents. As in the case of research time, reporting a dependent under 5 years old is associated with the largest declines. Second, gender differences in these forecasts appear attributable to differential changes associated with dependents. Reporting a 6- to 11-year-old dependent is associated with a 6.6% (s.e.=1.9) and 5.4% (s.e.=1.8) lower forecast of publication quantity and impact, respectively, but only for female scientists (see SI S5.3 for the field-level results).

We find that most of the same groups currently reporting the largest disruptions to research time also report the worst outlook for future publications. The correlations between reported change in research time and forecasted publication output are 0.337 for quantity (p-value < 0.001) and 0.214 for impact (p-value < 0.001). While understanding the relationships between time input and research output is beyond the scope of this study, we repeat the analysis, including the changes in reported time allocations to test if they moderate the effects we observe. We find that, while the post-Lasso regression coefficients associated with the selected demographic features generally become smaller, a statistically significant relationship remains in most cases even when conditioning on the (Lasso-selected) change in research time. This suggests the forecasted declines associated with reporting young dependents are not simply explained by the direct change in time spent on research (Figure S7).

We further investigate how these publication forecasts may depend on the expected duration of the COVID-19 pandemic by plotting the (randomized) expectation shown to the survey respondent against the estimated net effect of the pandemic (Figure 3C). A linear fit indicates that, for every 1 month that the pandemic continues past April 2020, scientists expect a 0.63% decrease in publication quantity (s.e.=0.23) and a 0.48% decrease in impact (s.e.=0.19) due to the pandemic. These marginal effects may appear small relative to the others documented in this paper, but it is important to note that they are on a similar scale as economic forecasts for the U.S. and Europe, which (as of May 2020) project economic declines in the range of 0.4-0.6% per month (5-7% for 2020)[33]. Still, these results could also reflect uncertainties or errors inherent to these forecasts, or strong personal beliefs about the timeline for the pandemic that are not easily swayed by the survey's suggestion.

**Discussion and concluding remarks**

Our results shed light on several important considerations for research institutions as they consider reopening plans and develop policies to address the pandemic's disruptions. The findings regarding the impact of childcare reveal a specific way in which the pandemic is impacting the scientific workforce. Indeed, "shelter-at-home" is not the same as "work-from-home" when dependents are also at home and need care. Because childcare is often difficult to observe and rarely considered in science policies (aside from parental leave immediately following birth or adoption), addressing this issue may be an uncharted but important new territory for science policy and decision makers. Furthermore, it suggests that unless adequate childcare services are available, researchers with young children may continue to be affected regardless of the reopening plans of institutions. And since the need to care for dependents is by no means unique to the scientific workforce, these results may also be relevant for other labor categories.

More broadly, many institutions have announced policy responses such as tenure clock extensions for junior faculty. Of 34 U.S. university policies we identified that provided some form of tenure extension due to the pandemic, 30 appeared to guarantee the extension for all faculty (see SI S5.5 for more). Institutions may favor such uniform policies for several reasons such as avoiding legal challenges. But given the heterogeneous effects of COVID-19 we identify, it raises further questions whether these uniform policies, while welcoming, may have unintended consequences and could exacerbate pre-existing inequalities[34].

While this paper focuses on quantifying the immediate impacts of the pandemic, circumstances will continue to evolve and there will likely be other notable impacts to the research enterprise. The heterogeneities we observe in our data may not converge, but instead may diverge further. For example, when research institutions begin the process of reopening, there may be different priorities for "bench sciences" versus work that involves human subjects or that requires travel to field sites. And research requiring international travel could be particularly delayed; all of which could lead to new productivity differences across certain groups of scientists. Furthermore, individuals with potential vulnerabilities to COVID-19 may prolong their social distancing beyond official guidelines. In particular, senior researchers may have incentives to continue avoiding in-person interactions[35], which historically facilitate mentoring and hands-on training of junior researchers.

The possibility of a resurgence of infections[36] suggests that institutions may anticipate a reinstatement of preventative measures such as social distancing. This possibility could direct focus toward research projects that can be more easily stopped and restarted. Funders seeking to support high-impact programs may have similar considerations, favoring proposals that appear more resilient to uncertain future scenarios.

Lastly, although we have focused on two of the denser geographic regions of scientific output in this study, the pandemic is having a substantial impact on research worldwide. In the coming years, researchers may be less willing or able to pursue positions outside of their home nation, which may deepen or alter global differences in scientific capacity. Future work expanding our understanding of how the pandemic is affecting researchers across different countries, at different institutions, and in different points of their life and career could provide valuable insights to more effectively protect and nurture the scientific enterprise. The strong heterogeneities we observe, and the likely development of new impacts in the coming months and years, both argue for a targeted and nuanced approach as the world-wide research enterprise rebuilds.

**Human research participants**

The study protocol has been approved by the Institutional Review Board (IRB) from Harvard University and Northwestern University. Informed consent was obtained from all participants.

**Acknowledgements**

We thank Alexandra Kesick for invaluable help. This work is supported by the Air Force Office of Scientific Research under award number FA9550-19-1-0354, National Science Foundation SBE 1829344, and the Alfred P. Sloan Foundation G-2019-12485 and G-2020-13873.

**Author contributions**

KRM, JGT, MCT, KRL, DW conceived the project, KRM, WYT, NC, JGT, MCT, KRL, DW designed the experiments, KRM, WYT, YY, NC collected data, KRM, WYT, YY performed empirical analyses, all authors collaboratively interpreted results, KRM, DW wrote the manuscript, all authors edited the manuscript.

**Competing interest declaration**

The authors declare no competing interests.

**Data availability**

Because of the sensitive nature of some of the variables collected, the IRB-approved protocol does not permit individual-level data to be made unrestricted and publicly available. Researchers interested in obtaining restricted, anonymized versions of this individual-level data should contact the authors to inquire about obtaining an IRB-approved institutional data sharing agreement.

**Code availability**

Code necessary to reproduce all plots and statistical analyses will be made freely available.

**Figures**

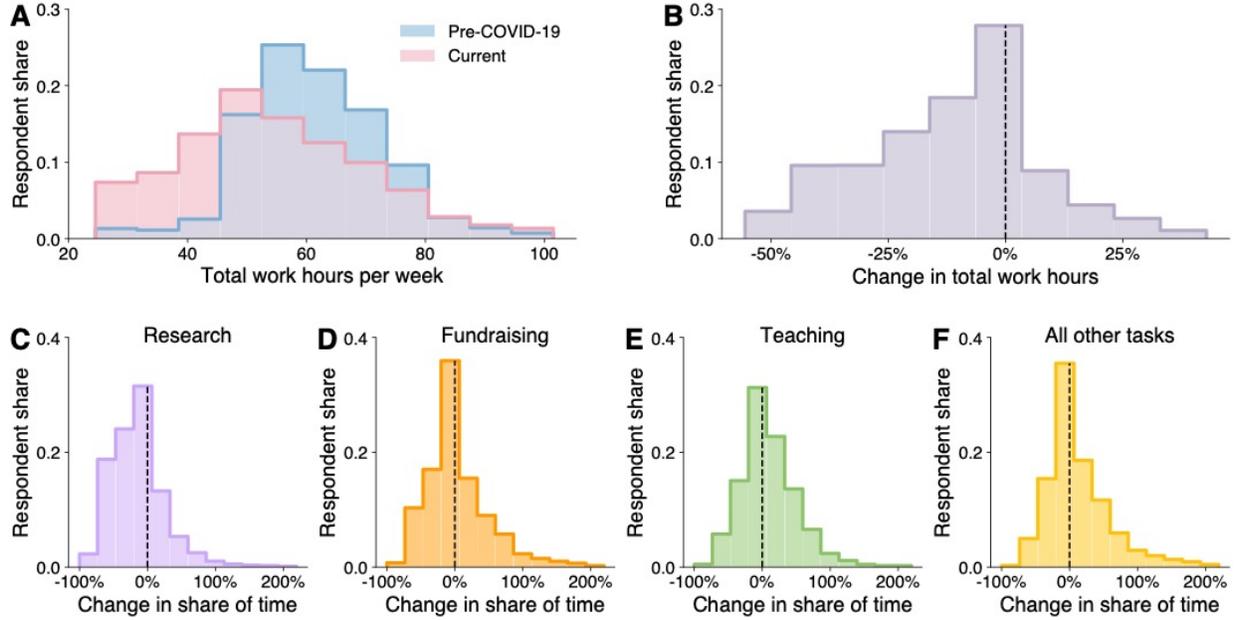

**Figure 1. Changes in the level and allocations of work time. A.** Distribution of total hours spent on work pre-pandemic and at the time of the survey. **B.** Distribution of changes in total work hours from pre-pandemic to time of survey. **C-F.** Distribution of percent changes in the share of work time allocated to research (C), fundraising (D), teaching (E), and all other tasks (F).

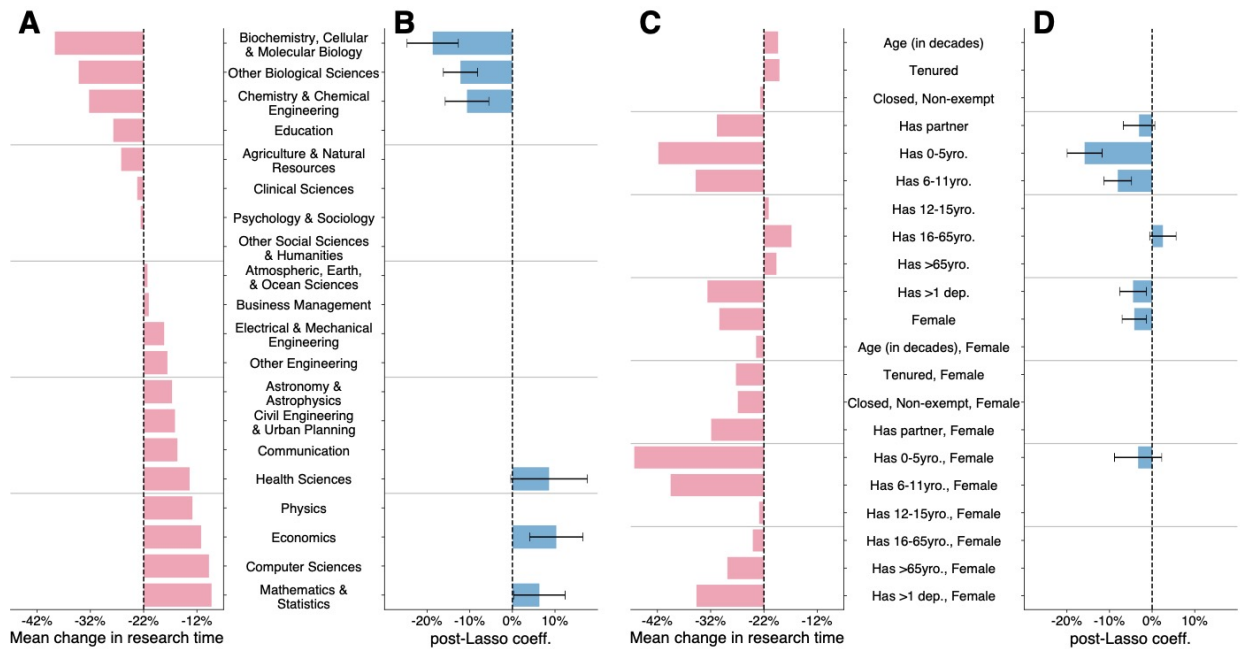

**Figure 2. Field and Group-level Changes in Research Time. A.** Field-level average changes in research time. The bars are centered around the sample-wide average change (approx. -22%). **B.** Post-Lasso regression coefficients for field indicators conditional on the flexible specification of demographic, career-stage, location, and pre-pandemic features. Error bars indicate 95% confidence intervals. If no bar is shown, the indicator was not selected in the Lasso. **C.** Group-level average changes in research time. Note that the figure is centered around the sample-wide average change (approx. -22%). **D.** Post-Lasso regression coefficients for group indicators conditional on the flexible specification of field, career-stage, location, and pre-pandemic features. Variable names with "Female" suffix indicate the variable is interacted with a female indicator. If no bar is shown, the indicator was not selected in the Lasso. Error bars indicate 95% confidence intervals.

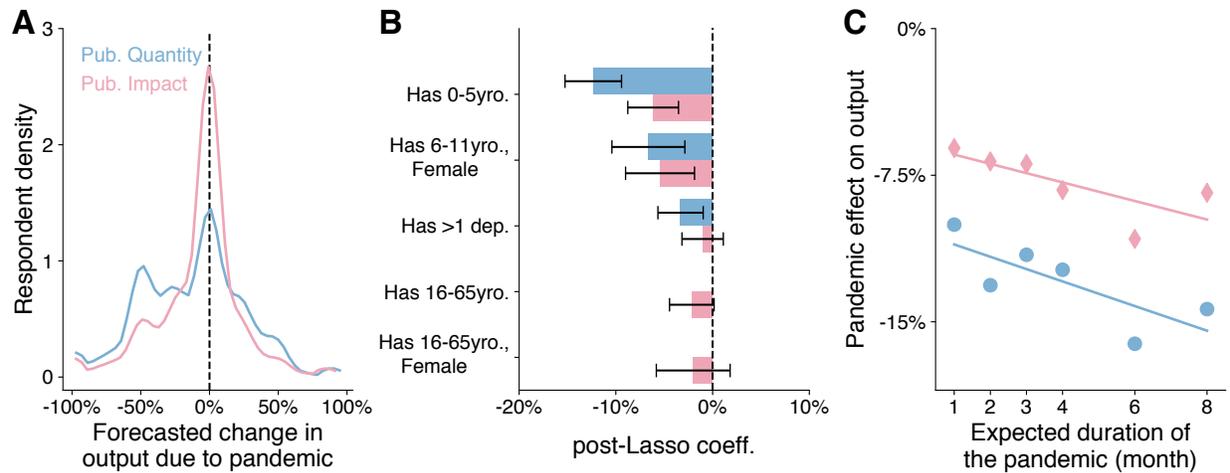

**Figure 3. Forecasted Changes in Publication Output. A.** Distributions of beliefs about the net pandemic effect on publication quantity (blue) and impact (red). **B.** Post-Lasso regression coefficients for selected demographic group indicators conditional on the flexible specification of field, career-stage, location, and pre-pandemic features. If no bar is shown, the indicator was not selected in the Lasso. Variables with "Female" suffix are interaction terms. Error bars indicate 95% confidence intervals. Figure S6 reports the results from a similar exercise focusing on field-level differences. We find the same three fields associated with the largest declines in research time—biochemistry, biology, and chemistry—also forecast the largest pandemic-induced declines in their publication output quantity, ceteris paribus. **C.** Average estimated changes in publication outputs per the randomized duration of pandemic respondents were asked to assume for their forecasts (either 1, 2, 3, 4, 6, or 8 months from the time of the survey, mid-April 2020).

# Supplementary Information for Quantifying the Immediate Effects of the COVID-19 Pandemic on Scientists

Kyle R. Myers[1,2*], Wei Yang Tham[1,2], Yian Yin[3,4,5], Nina Cohodes[2,6], Jerry G. Thursby[2,7], Marie C. Thursby[2,8], Peter E. Schiffer[9], Joseph T. Walsh[5,10], Karim R. Lakhani[1,2,6], Dashun Wang[3,4,5,11*]

[1]Harvard Business School, Harvard University, Boston, MA, USA
[2]Laboratory for Innovation Science at Harvard, Harvard University, Boston, MA, USA
[3]Center for Science of Science and Innovation, Northwestern University, Evanston, IL, USA
[4]Northwestern Institute on Complex Systems, Northwestern University, Evanston, IL, USA
[5]McCormick School of Engineering, Northwestern University, Evanston, IL, USA
[6]Institute for Quantitative Social Science, Harvard University, Boston, MA, USA
[7]TyGeron Institute, Nashville, TN, USA
[8]Scheller College of Business, Georgia Institute of Technology, Atlanta, GA, USA
[9]Department of Applied Physics and Department of Physics, Yale University, New Haven, CT, USA
[10]University of Illinois System, Urbana, IL, USA
[11]Kellogg School of Management, Northwestern University, Evanston, IL, USA
[*]Correspondence to: kmyers@hbs.edu, dashun.wang@northwestern.edu

**Table of Contents**

# S1 Survey Sampling & Recruitment

## S1.1 Web of Science corresponding authors

To compile a large, plausibly random list of active scientists, we leverage the Web of Science (WoS) publication database. The WoS database is useful for two reasons: (1) it is one of the most authoritative citation corpuses available[1] and has been widely used in recent science of science studies[2–4]; (2) among other large-scale publication datasets, WoS is the only one, to our knowledge, with systematic coverage of corresponding author email addresses.

We are primarily interested in active scientists residing in the U.S. and Europe. We start from 21 million WoS papers published in the last decade (2010-2019). In an attempt to focus on scientists likely to still be active and in a more stable research position, we link the data to journal impact factor information (WoS Journal Citation Reports), and exclude papers published in journals in the bottom 25% of the impact factor distribution for its WoS-designated category. We use the journal impact factor calculated for the year of publication, and for papers published in 2019, we use the latest version (2018). We then extract all author email addresses associated with papers. For each email address in this list, we consider it as a potential participant if: (1) it is associated with at least two papers in the ten-year period, and (2) the most recent country of residence, defined by the first affiliation of the most recent paper, is in the U.S. or Europe.

We have approximately 2.5 million unique email addresses after filtering, with about 521,000 in the U.S. and 938,000 in Europe. We then randomly shuffled the two lists separately and sampled roughly 280,000 email addresses from the U.S. and 200,000 from Europe. We oversampled the U.S. as a part of a broader outreach strategy underlying this and other research projects.

## S1.2 Participant recruitment

We recruited participants by sending them email invitations through with the following text:

> *We need your help to shed light on how the coronavirus pandemic is affecting scientists like you. Please take a brief moment to complete this short five-minute survey as part of a research study. Your responses will help scientists and policymakers understand and respond to this rapidly evolving situation.*
>
> ***Follow this link to the Survey****:*
> *[link]*
> ***Or, copy and paste the URL below into your internet browser****:*
> *[link]*
>
> *Upon completion, you can choose charities to receive a donation on your behalf, and you may have the chance of winning a $100 gift card. Please feel free to forward this survey to any other scientists you know (e.g., professors, post-doctoral researchers, graduate students). We need everyone's input to fully understand the breadth of how science is currently changing.*
>
> *Thank you for your time,*

*Kyle Myers, Ph.D. & Karim Lakhani, Ph.D.*
*Laboratory for Innovation Science at Harvard*

*Dashun Wang, Ph.D.*
*Northwestern Kellogg Center for Science of Science & Innovation (CSSI)*

## S2 Survey Instrument

### S2.1 Survey questions

The survey includes questions on professional information (position type, institution type, fields of study, type of research, tenure status), demographic information (age, gender, cohabitation, dependents, citizenship), time allocation (time spent on different activities before and after the pandemic outbreak) and predicted changes in future publication and funding. Below are the excerpted questions underlying the variables used in our analyses. We did not require respondents to answer any of the demographic questions regarding gender, age, dependents, cohabitation, or citizenship.

*Q. Which of the following best describes your current position?*
- *Faculty or Principal Investigator | Post-doctoral researcher | Graduate student in a doctoral program | Retired scientist no longer engaged in research | Other*

*Q. Which of the following best describes your field of study?*
- *General: [list of 20 fields]*

*Q. Which of the following best describes the institution you are primarily affiliated with?*
- *University or college | Non-profit research organization | Government or public agency | For-profit firm | Other*

*Q. Please answer the following:*
- *Is your institution physically closed to non-essential personnel?*
    - *Yes | No | Not relevant*
- *Are you exempt from the closure and allowed to travel to your work site(s)?*
    - *Yes | No | Not relevant*
- *Do you have tenure?*
    - *Yes | No | Not relevant*

*Q. Gender:*
- *Male | Female | Other | Prefer not to say*

*Q. Age:*
- *Under 20 | 20-24 | 25-29 … 75-79 | 80 or older | Prefer not to say*

*Q. Number of dependents of any age you care for:*
- *0 | 1 | 2 | 3 or more | Prefer not to say*

*Q. In what age group(s) are your dependents? Note. You may select multiple*
- *0-2 years old | 3-5 years old | 6-11 years old | 12-18 years old | 18-65 years old | Over 65 years old*

*Q. Cohabitation status:*
- *I reside with a partner, spouse, or significant other | I reside with friends | I reside by myself | Other | Prefer not to say*

*Q. Are you a U.S. Citizen?*
- *Yes | No | Prefer not to say*

*Q. Before this coronavirus pandemic, about how many hours per week did you work on anything related to your job? (e.g., researching, teaching, writing)*

- *14-21 hours per week (avg. 2-3 hours every day) | 21-28 hours per week (avg. 3-4 hours every day) | … | 77-84 hours per week (avg. 11-12 hours every day) | More than 84 hours per week (avg. 12 hours or more every day)*

*Q. Currently, about how many hours per week are you working? (e.g., researching, teaching, writing)*

- *14-21 hours per week (avg. 2-3 hours every day) | 21-28 hours per week (avg. 3-4 hours every day) | … | 77-84 hours per week (avg. 11-12 hours every day) | More than 84 hours per week (avg. 12 hours or more every day)*

*Q. Before this coronavirus pandemic, how did you spend your work hours?*
*Research: \_\_\_\_\_\_\_*
*Teaching, advising, or mentoring: \_\_\_\_\_\_\_*
*Grant writing or other fund-raising: \_\_\_\_\_\_\_*
*Administration, service, or other (e.g., editorial duties, peer review): \_\_\_\_\_\_\_*
*Total: \_\_\_\_\_\_\_\_*

*Q. Currently, how are you spending the hours you do work?*
*Research: \_\_\_\_\_\_\_*
*Teaching, advising, or mentoring: \_\_\_\_\_\_\_*
*Grant writing or other fund-raising: \_\_\_\_\_\_\_*
*Administration, service, or other (e.g., editorial duties, peer review): \_\_\_\_\_\_\_*
*Total: \_\_\_\_\_\_\_\_*

*Q. Previously in 2018 and 2019, approximately how much research funding did you oversee or manage per year? Note: Ignore overhead or indirect costs; focusing only on funds used directly for research regardless of source (e.g., home institution, federal grant).*
*Approximately:*
- *<10,000 $/year | 10,000-20,000 $/year | 20,000-50,000 $/year | 50,000-100,000 $/year | 100,000-200,000 $/year | 200,000-500,000 $/year | 500,000-1,000,000 $/year | 1,000,000-2,000,000 $/year | >2,000,000 $/year*

*For the following, please consider your "research publications" as all of your publications that focus on a research question. (e.g., journal articles, conference proceedings, patents, books. Ignore commentary, editorials, etc.)*

*Q. Assuming this pandemic lasts for another* [1, 2, 3, 4, 6, 8] *months (until* [May, June, July, August, October, December]*), how do you think the quantity and impact of your research publications will change in 2020 and 2021 compared to 2018 and 2019? (i.e., what will be the percent change?)*
- *Quantity (i.e., number):* [slider scale from -100% to +50% in 10% increments]
- *Impact (i.e., quality influence):* [slider scale from -100% to +50% in 10% increments]

### S2.2 Field definitions

We build on field classifications used in national surveys such as the U.S. Survey of Doctorate Recipients (SDR) to categorize fields in our survey, aggregating to ensure sufficient sample sizes within each field. The notable additions we make to the fields used in these other surveys are to include: Business Management, Education, Communication, and Clinical Sciences. These fields reflect major schools at most universities and/or did not immediately map to some of the default fields used in the SDR (i.e., the "Health Sciences" field in SDR does not include medical specialties).

## S3 Sampling Approach

### S3.1 Distribution and Basic statistics

Out of a total of 480,000 emails sent, approximately 64,000 emails were directly bounced either due to incorrect spelling in the WoS data or the termination of the email account. In hopes of soliciting a larger sample, we also undertook snowball sampling by encouraging respondents to share the survey with their colleagues as well.

Overall 9,968 individuals entered the survey and 8,447 continued past the consent stage. Of those that did not, 412 were not an active scientist, post-doc, or graduate student and thus not within our population of interest, 81 did not consent, and 1028 did not make any consent choice. When a respondent continued past the consent stage, we asked them to report the type of role they were in. Out of the 8,447 consenting responses, there 5,728 responses from faculty or principal investigators (PIs), 1,023 responses from post-doctoral researchers, 701 from graduate students in a doctoral program, and 52 from retired scientists. 551 of the remaining respondents were some other type of position and another 392 did not report their position. This yields an estimate of a response rate of approximately 1.6%. First, our low response rate may reflect the disruptive nature of the pandemic, but it also raises concerns for generalizability of our results. However, after we received feedback from the initial distribution that many individuals had received the email in their "junk" folder, we became concerned with our distribution being automatically flagged as spam. Based on spot-checking of five individuals that we ex-post identified as being randomly selected by our sample, and who we had professional relationships with, found that in four of the five cases the recruitment email had been flagged as spam. We know of no systematic way of estimating the true spam-flagging rate (nor how to avoid these spam filters when using email distributions at this scale) without using high-end, commercial-grade products.

Additionally, as with any opt-in survey, there may be correlations between which scientists opt-in and their experiences about which they want to report. For example, scientists who felt strongly about sharing their situation, whether they experienced large positive or negative changes, may be more likely to respond, which would increase the heterogeneity of the sample. Furthermore, there may also be non-negligible gender differences that arise not due to actual differences in outcomes but due to differences in reporting known to occur across genders[5–9].

For our analyses, we focus entirely on responses from the sample of faculty/PIs. From the full sample of PIs, we retain respondents who reported working for a "University or college", "Non-profit research organization", "Government or public agency", or "Other", and excluding 87 responses from individuals who report to work for a "For-profit firm". We also restrict the sample to respondents whose IP address originated from the United States or Europe (dropping 1,049 responses from elsewhere).

We then drop observations that have missing data for any of the variables used in our analyses: 26 responses do not report their time allocations, 74 do not report their age, 10 do not report the type of institution they work at, and 114 do not report their field of study. Altogether, this amounts to dropping 187 observations. Given the relatively small subset of our sample dropped due to missing data, we do not impute missing variables as this introduces unnecessary noise[10].

The summary statistics for the final sample used in the analyses are reported in Figure S1 and the geographic distribution of respondents is shown in Figure S2.

## S3.2 Comparison to SDR

To estimate the generalizability of our respondent sample, we use the public microdata from The Survey of Doctorate Recipients (SDR) as the best sample estimates of the population of principal investigators in the U.S. The SDR is conducted by the National Center for Science and Engineering Statistics within the National Science Foundation, sampling from individuals who have earned a science, engineering, or health doctorate degree from a U.S. academic institution and are less than 76 years of age. The survey is conducted every two years, and we use the latest data available (2017 cycle). For this comparison, we focus only on university faculty in both our survey and the SDR. We also constrain our sample to only include fields of study with a clear mapping to the SDR categories. The SDR focuses only on researchers with Ph.D.-type degrees, and so it does not capture researchers with other degrees still actively engaged in research (i.e., researchers with only M.D.s). This means we exclude "architecture and design," "business management," "medicine," "education," "humanities," and "law and legal studies." Figure S2 compares respondents between our sample and the SDR sample.

Figure S3a illustrates differences on demographics and career-stage features, including raw differences as well as those adjusted by field. We find only a small difference in age and no difference in partner status. Our survey oversamples on female scientists, those with children, and untenured faculty. These differences persist after conditioning on the scientist's reported field. That we ultimately find female scientists and those with young dependents to report the largest disruptions suggests that these individuals may have been more likely to respond to the survey in order to report their circumstances. The geographic distributions are relatively similar, with slight oversampling of west and undersampling of south. Lastly, we find a significant but small oversampling of U.S. citizens.

We also compare the distribution of research fields (Fig. S3.b). Overall the distributions are relatively similar. We appear to oversample most significantly on "atmospheric, earth, and ocean sciences" and "other social sciences." While we undersample most significantly on the biological sciences, "mathematics and statistics," and "electrical and mechanical engineering". There does not seem to be a clear pattern with these field-level differences, as we undersample fields that ultimately report being across the spectrum of disruptions (i.e., mathematics and statistics reports some of the smallest disruptions, and the biological sciences are amongst the most disrupted).

## S4 Sampling Approach

### S4.1 Lasso selection and post-Lasso regression

The unconditional changes reported by each group of scientists is informative of how the pandemic affected researchers overall. But it does not allow us to infer whether groups reporting larger or smaller disruptions are doing so for reasons inherent to that group (i.e., the nature of work in certain fields, or the demands of home life unique to certain individuals) or because the individuals that select into that group tend to also be disrupted for unrelated reasons. This motivates a multivariate regression analysis to explore whether changes associated with a group of individuals change after conditioning on other observables. However, selecting which of an available set of covariates (or transformations thereof) to include in a regression is notoriously challenging. The Lasso method provides a data-driven approach to this selection problem by excluding covariates from the regression that do not improve the fit of the model[11,12].

When using the Lasso, our general approach is to include a vector of indicator variables for the fields or demographic/career groups of interest, along with an additional set of controls. When focusing on differences across fields, we include the demographic/career variables in the control set, and vice versa. The control variables common to all Lasso-based analyses are: pre-pandemic level of time allocations and totals, pre-pandemic share of time allocations, pre-pandemic funding estimate, and indicators for the type of institution (academic, non-profit, government, or other) and the location (state if in U.S., country if in Europe). To make minimal assumptions about the functional form of the control variables, we conduct the following transformations to expand the set of controls: for all continuous variables we use inverse hyperbolic sine (which approximates a logarithmic transformation while allowing zeros), square and cubic transformations, and we interact all indicator variables with the linear versions of the continuous variables.

We perform the Lasso using the lasso linear package in Stata 16 © software. We use the defaults for constructing initial guesses, tuning parameters, number of folds (ten), and stopping criteria. We use the two-step cross-validation "adaptive" Lasso model where an initial instance of the algorithm is used to make a first selection of variables, and then a second instance occurs using only variables selected in the first instance. The variables selected after this second run are then used in a standard post-Lasso OLS regression with heteroskedastic robust standard errors.

### S4.2 Estimating net pandemic effects on publication output

We are interested in the effect of the COVID-19 pandemic on research output. As an initial estimate of what this effect could be, we asked respondents to forecast how their research output in 2020 and 2021 will compare to their prior output in 2018 and 2019. This framing was chosen for its simplicity; however, it does not provide a direct estimate of the pandemic effect. For this effect, we could have asked how the respondent expects their output to be in 2020 and 2021 compared to what they would otherwise expect their output to have been in 2020 and 2021 had the pandemic not occurred. Clearly, this is more complicated. But since we chose the simpler framing, we must account for some underlying factors before arriving at figures closer to what scientists think the effect of the pandemic will be (or our estimates thereof).

These raw year-to-year forecasted changes in publication outputs will be influenced by four major factors: (1) changes due to the pandemic to date; (2) anticipated future changes due to the pandemic; (3) the respondent's expectations about how long the pandemic will last; and (4) regular trends in the evolution of publication output across different individuals and fields (e.g., if female scientists have continually been increasing their number of publications produced each year, then in the absence of the pandemic we might expect this trend to continue into the near future). Again, we are primarily interested in (1) and (2). To overcome (3), we randomly assign respondents to make forecasts for one of 6 possible scenarios where they were to take as given the duration of the pandemic to be either 1, 2, 3, 4, 6, or 8 months from the time of the survey. In some analyses, we condition on this variable to control for variation due to perceptions about the length of the pandemic. In others, we explore the effect of these different perceptions directly to infer how scientists perceive disruptions may evolve as the pandemic does or does not continue to persist.

With respect to the issue of differential trends across individuals and fields, we first note that the time scale we are concerned with (approx. 2 years) is small enough that we expect the majority of individuals to not change in terms of their observables. This is because all of our time-dependent observables used in the analyses are based on groupings of 5 years. Still, to more quantitatively address this issue, we use historical data and another Lasso-based regression model to project scientists' publication output in 2020 and 2021, using their observable features from the survey and publication data since 2010. Our assumption is that these projections can approximate what scientists would have forecasted in the absence of the pandemic--they provide a crude counterfactual. Given the short timeframes involved, and the rich observable data we possess, we hypothesize that the room for significant biases or deviations are small relative to the across-individual variation.

Due to data quality limitations, we are only able to connect 56% of respondents to their publication records, but a comparison of observables indicates that there are no meaningful differences between those scientists connected to their publication record, and those not (See Figure S4). Since we observe the variables used in these projections for all respondents, we can project out trends for all scientists in our sample.

While the measurement of publication quantity is straightforward, the measurement of quality or, as it was asked in the survey, "impact" is not. Following a long line of science of science research[13], we use citation counts as the best available proxy for quality. We follow the state of the art in terms of adjusting and counting these citations in a manner that does not conflate across-field differences[14].

The Lasso-based projection proceeds as follows. First, we demean the publication measures at the year level. This is because we do not want to attribute aggregate year-to-year variations across the entire sample to actual changes in net output, since these fluctuations can very plausibly be linked to changes in the Web of Science (WoS) coverage over time, and we are much more concerned with differential trends amongst different fields and/or different individuals. Next, we use the Lasso to select which of the observables are the best predictors of publication counts and citations. The major difference between this Lasso-based approach and the others used in this paper is that, here, we interact all observables with flexible time trends (i.e., squared, cubic, and inverse hyperbolic sine transformations of the year variable) to allow differential trends across groups. Finally, we project out these expected output measures as a function of the selected covariates and their corresponding coefficients from a post-LASSO OLS regression. Importantly, we project out of sample just two years so that we have estimates of the counterfactual trends for 2020 and 2021. With these estimates of respondents' counterfactual forecasts in hand, we then simply subtract them from scientists actual reported forecasts to arrive at our estimate of scientists' forecast of the "net effect" of the pandemic. Figure S3 plots the distributions of the unadjusted forecasts and these net effects for both the quantity and impact measures. The adjustment does not substantially change the distribution, but we are more confident in these estimates as "pandemic effects" for the aforementioned reasons.

## S5 Additional Results

### S5.1 Correlations in time spent on different tasks

Figure S5 plots the reported changes in research time (y-axes) against the reported changes in time allocated to the other three task categories (x-axes). The figures are binned scatterplots, and linear fits of the data suggest that research may be a substitute for the other categories. A 10% increase in fundraising, teaching, or all other tasks is associated with a decline in research by 1.4% (s.e.=0.14), 4.6 % (s.e.=0.16), and 3.2% (s.e.=0.13), respectively. We lack exogenous variation in the data that can clearly shift the time allocated to one (or a subset) of tasks, so we cannot identify the extent to which these correlations reflect actual substitution patterns or unobserved factors. Though the magnitudes and precision of these relationships suggests further investigations are certainly warranted to better understand how scientists allocate their time.

### S5.2 Predictors of changes in non-research tasks by groups

Figure S6a and S6b replicate Figures 2b and 2d from the main text, respectively, instead using each of the other three task categories for the dependent variable. For the analysis focused on fields (Fig. S6a), no clear patterns emerge with respect to changes in time spent fundraising or teaching. Reported time changes in teaching may be due to a combination of reasons. First, during the pandemic, the demand for these activities is likely relatively stable (e.g., most academic institutions have moved classes online, but there are few reports of suspension of classes); and second, impacts due to the transition to online teaching may have taken place earlier, hence not captured by our survey. There is evidence that clinical science and biochemists are spending an increasing amount of time on the "all other tasks" category, which could plausibly be due to a redirection of effort directly towards pandemic-related (non-research) work.

For the analysis focused on demographic groups (Fig. S6b), we find that scientists reporting a dependent under 5 years old tend to also report larger declines across all task categories. This result is consistent with an unsurprising hypothesis that these dependents require care that leads scientists to decrease their total work hours. The fact that there does not appear to be any substitution away from research towards these other categories for these specific individuals with young dependents suggests the association is driven by factors inherent to having a dependent at home, and not that these individuals also tend to select alternative work structures that has them performing less research and more of other tasks.

### S5.3 Additional Publication Forecast Results

Figure S7 recreates Figure 3b from the main text, but using the field-level Lasso approach. Forecasted changes in output are almost entirely confined to publication quantity (as opposed to impact), with the same fields of biology and chemistry that reported the largest declines in research time also forecasting the largest declines in publication output, here in the range of 4-10% relative to what would have been expected otherwise. Notably, some fields expect to publish more because of the pandemic, again highlighting the heterogenous experiences scientists are having due to the pandemic.

Figure S8 recreates Figure 3b from the main text, but while including the reported changes in time allocated to each of the four task categories (in addition to the pre-pandemic reported time allocations as before). Again, we find a similar set of dependent-related variables to be most predictive of forecasted publication changes, even though the reported change in research time is also selected as relevant by the Lasso. For comparison, the forecasted disruption associated with a dependent under 5 years old (7.04% decline expected publication count) is approximately the same magnitude as the implied effect associated with a 26% decrease in research time.

### S5.5 University Faculty Tenure Clock Policies

Using internet searches, we attempted to identify university-level tenure clock extension policies put in place as a result of the COVID-19 pandemic. While not a comprehensive list, we identified policies for 34 universities, encompassing both public and private, small and large institutions. Of the 34 universities, 17 have automatically applied a tenure clock extension to all faculty, with individuals having the ability to opt out[15–31]; 13 require applications but are automatically approved[32–44]. Four universities have not established unilateral policies[45–48]. Instead, they have either created a separate application process or added COVID-19-related impact to the list of reasons a faculty member may apply for an extension.

# S6 Supplementary Tables

| | N Obs. | Mean | Median | S.D. | Mean, with pubs. | Mean, miss. pubs. | t-stat |
|---|---|---|---|---|---|---|---|
| Research HRs. per week, pre-pandemic | 4535 | 23.63 | 22.40 | 10.82 | 23.46 | 23.86 | 1.26 |
| Teaching HRs. per week, pre-pandemic | 4535 | 16.39 | 15.75 | 9.21 | 16.51 | 16.23 | -1.03 |
| Fundraising HRs. per week, pre-pandemic | 4535 | 8.58 | 7.70 | 5.46 | 8.65 | 8.49 | -0.97 |
| All other task HRs. per week, pre-pandemic | 4535 | 12.80 | 10.50 | 8.29 | 12.85 | 12.74 | -0.47 |
| Total Work HRs. per week, pre-pandemic | 4535 | 61.41 | 63.00 | 11.72 | 61.47 | 61.32 | -0.44 |
| Research HRs. per week, current | 4535 | 17.85 | 14.70 | 12.65 | 17.65 | 18.11 | 1.20 |
| Teaching HRs. per week, current | 4535 | 16.08 | 13.30 | 11.93 | 16.17 | 15.96 | -0.59 |
| Fundraising HRs. per week, current | 4535 | 7.79 | 4.90 | 6.92 | 7.84 | 7.72 | -0.61 |
| All other task HRs. per week, current | 4535 | 12.72 | 9.80 | 9.87 | 12.88 | 12.52 | -1.23 |
| Total Work HRs. per week, current | 4535 | 54.44 | 56.00 | 15.87 | 54.54 | 54.30 | -0.51 |
| Age | 4535 | 47.28 | 45.00 | 11.39 | 47.16 | 47.43 | 0.80 |
| Has Tenure | 4535 | 0.60 | 1.00 | 0.49 | 0.60 | 0.60 | -0.05 |
| Female | 4535 | 0.41 | 0.00 | 0.49 | 0.41 | 0.41 | -0.12 |
| Lives with Partner | 4535 | 0.84 | 1.00 | 0.37 | 0.85 | 0.82 | -2.69 |
| U.S. Citizen | 4535 | 0.57 | 1.00 | 0.50 | 0.58 | 0.55 | -2.04 |
| Has Dependent, 0-5yro. | 4535 | 0.19 | 0.00 | 0.39 | 0.20 | 0.19 | -0.72 |
| Has Dependent, 6-11yro. | 4535 | 0.23 | 0.00 | 0.42 | 0.24 | 0.21 | -1.80 |
| Has Dependent, 12-15yro. | 4535 | 0.15 | 0.00 | 0.35 | 0.15 | 0.14 | -0.66 |
| Has Dependent, 16-65yro. | 4535 | 0.29 | 0.00 | 0.46 | 0.29 | 0.30 | 0.88 |
| Has Dependent, >65yro. | 4535 | 0.05 | 0.00 | 0.22 | 0.05 | 0.05 | 0.13 |
| Has >1 Dependent | 4535 | 0.41 | 0.00 | 0.49 | 0.41 | 0.40 | -0.52 |
| Inst. closed, non-exempt | 4535 | 0.649 | 1.000 | 0.477 | 0.645 | 0.656 | 0.765 |
| At University/College | 4535 | 0.85 | 1.00 | 0.36 | 0.86 | 0.82 | -3.51 |
| In U.S. (vs. Europe) | 4535 | 0.64 | 1.00 | 0.48 | 0.65 | 0.63 | -1.42 |
| Approximate Research Funding, pre-pandemic | 4535 | $ 235,398 | $ 35,000 | $ 499,425 | $ 240,701 | $ 228,556 | -0.81 |
| Pub. Quantity Forecast (`20-21 vs. `18-19) | 4535 | -12.43% | 0.00% | 33.05% | -11.69% | -13.37% | -1.70 |
| Pub. Impact Forecast (`20-21 vs. `18-19) | 4535 | -7.38% | 0.00% | 24.84% | -6.98% | -7.90% | -1.23 |
| Pub. Quantity (Number) since 2010, if matched | 2555 | 6.88 | 4.00 | 7.92 | | | |
| Pub. Impact (Eucl. Citations) since 2010, if matched | 2555 | 7.84 | 3.85 | 13.40 | | | |

**Table S1. Summary Statistics.** Summary statistics for the main survey sample. "Mean, with pubs." And "Mean, miss. pubs." report the averages for the sub-samples that can and cannot be connected to their publication record in WoS, respectively. The "t-stat" column reports the t-statistic from a test of mean differences between these two sub-samples. The two WoS-based variables are "Pub. Quantity (Number) since 2010" (the sum of the author's number of publications in the WoS record), and "Pub. Impact (Eucl. Citations) since 2010" (the field-demeaned euclidean sum of citations to the author's publications in the WoS record[14]).

# S7 Supplementary Figures

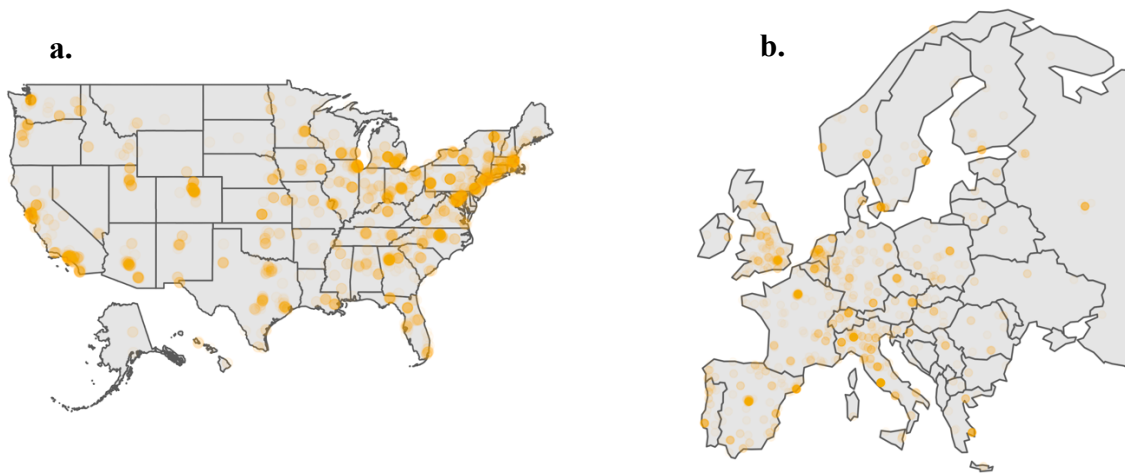

**Figure S1. Geographic distribution.** Plots respondent locations in U.S. (**a.**) and Europe (**b.**), aggregated to preserve anonymity.

**a.**

|  | SDR Sample | Survey Sample | Diff. | Diff., field adjusted |
|---|---|---|---|---|
| Age | 45.67 | 44.77 | -0.900*** | -1.054*** |
| Is Female | 0.37 | 0.48 | 0.111*** | 0.116*** |
| Has Dependent, 0-5yro. | 0.10 | 0.14 | 0.045*** | .047*** |
| Has Dependent, 6-11yro. | 0.06 | 0.10 | 0.036*** | 0.035*** |
| Has Dependent, 12-18yro. | 0.10 | 0.20 | 0.091*** | 0.088*** |
| Has Partner | 0.84 | 0.84 | -0.003 | -0.003 |
| Has Tenure | 0.67 | 0.61 | -0.061*** | -0.061*** |
| Located in Midwest | 0.22 | 0.23 | 0.009 | 0.017 |
| Located in West | 0.22 | 0.28 | 0.066*** | 0.054*** |
| Located in South | 0.33 | 0.24 | -0.091*** | -0.091*** |
| Located in Northeast | 0.22 | 0.24 | 0.015 | 0.020* |
| U.S. Citizen | 0.84 | 0.86 | 0.018** | 0.011 |

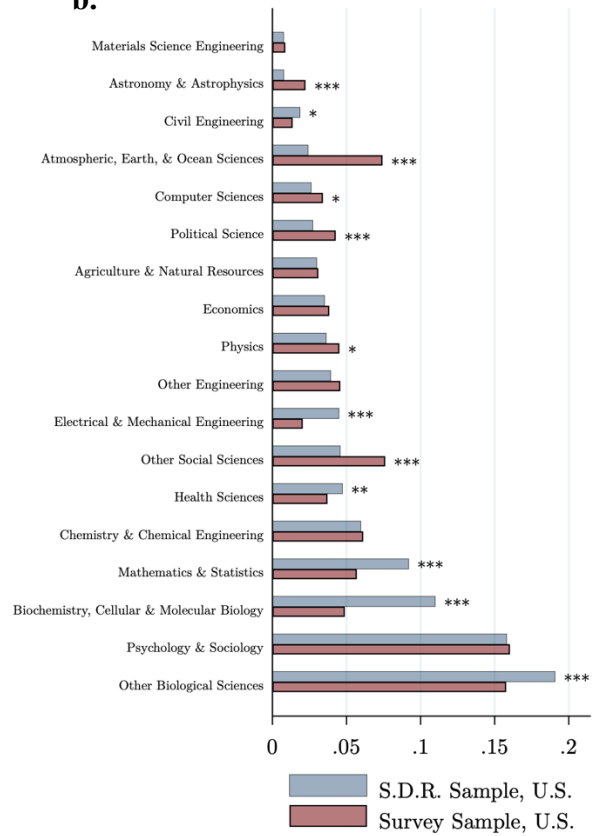

**Figure S2. Comparison to U.S. University-based SDR respondents.** Summary statistics for demographic variables and fields common to both our survey and the U.S. Survey of Doctoral Recipients (SDR). All comparisons are based on U.S.-located faculty or PIs at universities or colleges that report affiliation with a field of study present in both surveys (note: all fields present in the SDR are present in our survey, but not vice versa). **a.** Describes the sample averages for both samples and the mean differences in both the raw data ("Diff.") and after adjusting for the different composition of fields in each sample ("Diff., field adjusted") **b.** Plots the share of respondents in each sample that affiliate with each of the fields common to both surveys. (*** $p<0.01$; **$p<0.05$; *$p<0.1$)

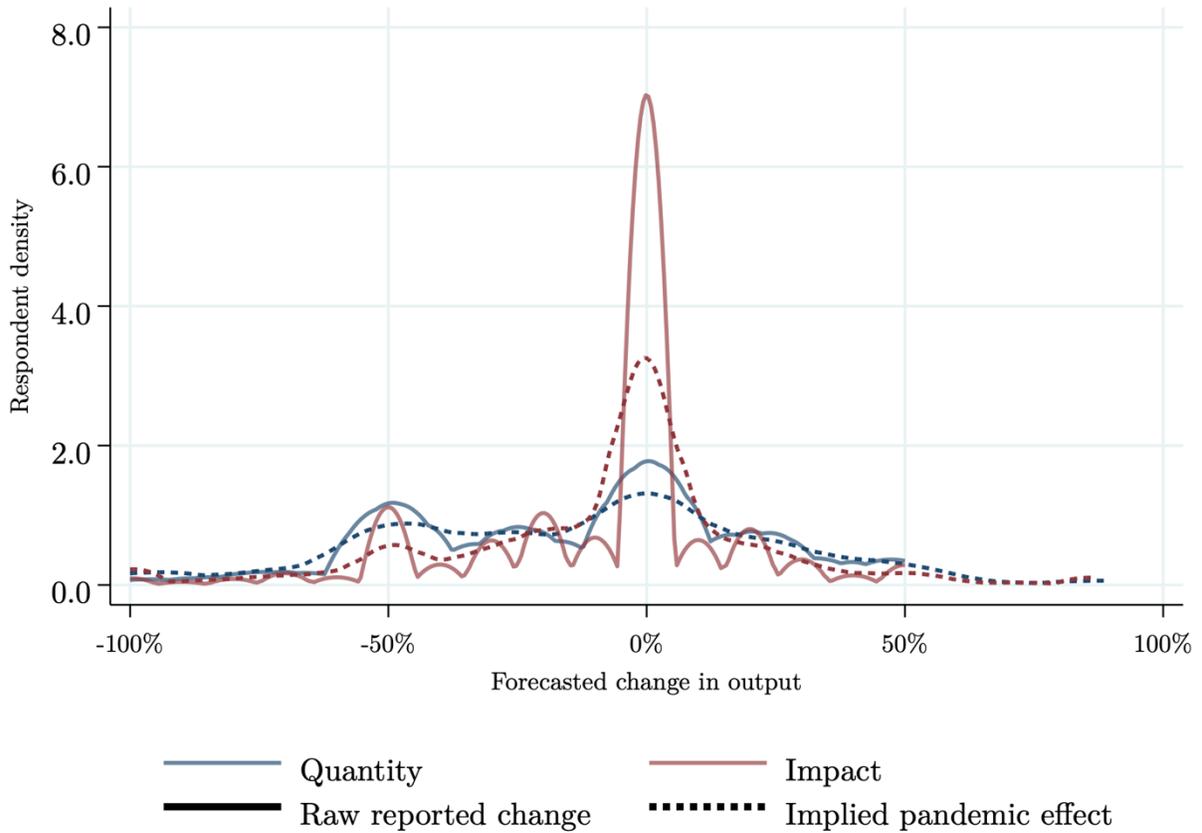

**Figure S3. Publication changes, raw and inferred pandemic effects.** Plots the distribution of changes to publication output. Blue lines indicate publication quantity, red lines indicate impact. Solid lines indicate the raw responses from the survey (which asked only about changes in publication output from 2018-19 to 2020-21), and dashed lines indicate our estimates of the implied effect due to the COVID-19 pandemic based on the removal of group-specific trends in publication output. See the Methodology section 2 for more.

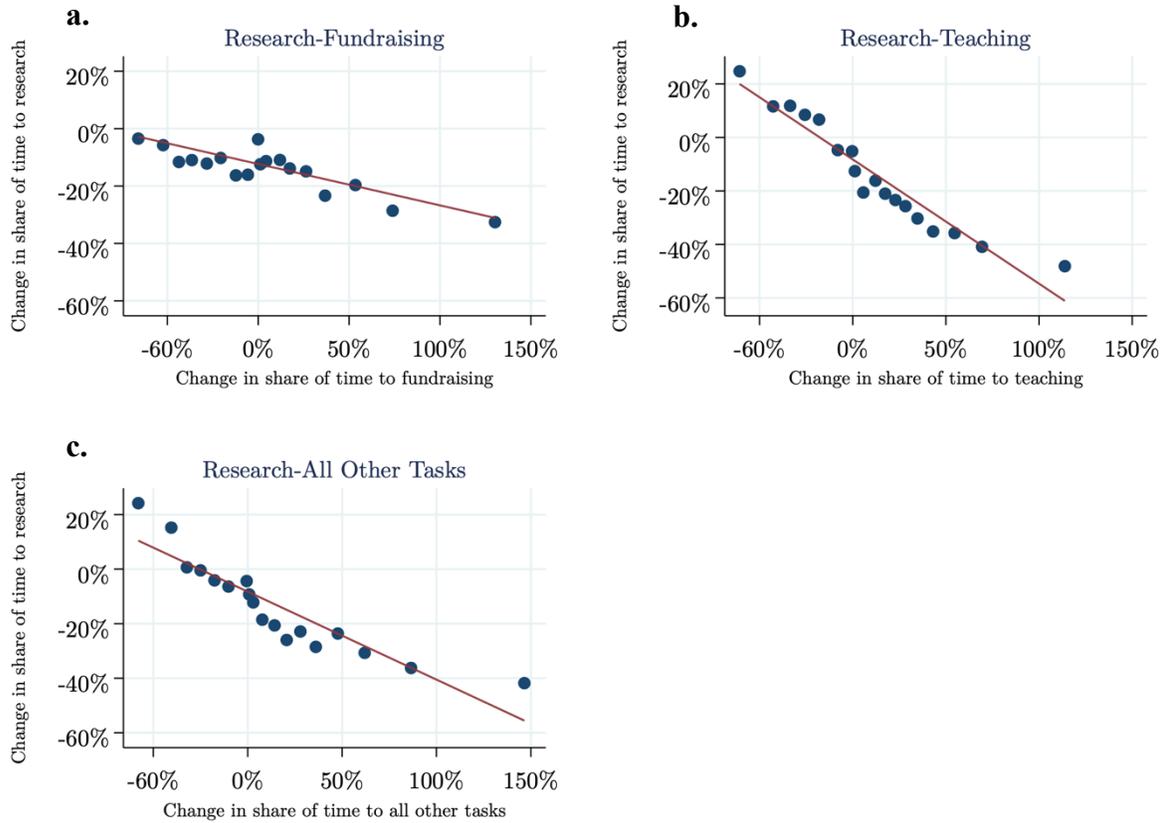

**Figure S4. Correlation between changes in task allocations, by category.** Binned scatterplots (where each of the 18 dots represent approximately 5.5% of observations) of the relationship between the change in reported time spent on research versus fundraising (**a.**), teaching (**b.**), or the "all other task" residual category (**c.**). Linear fit lines are also plotted.

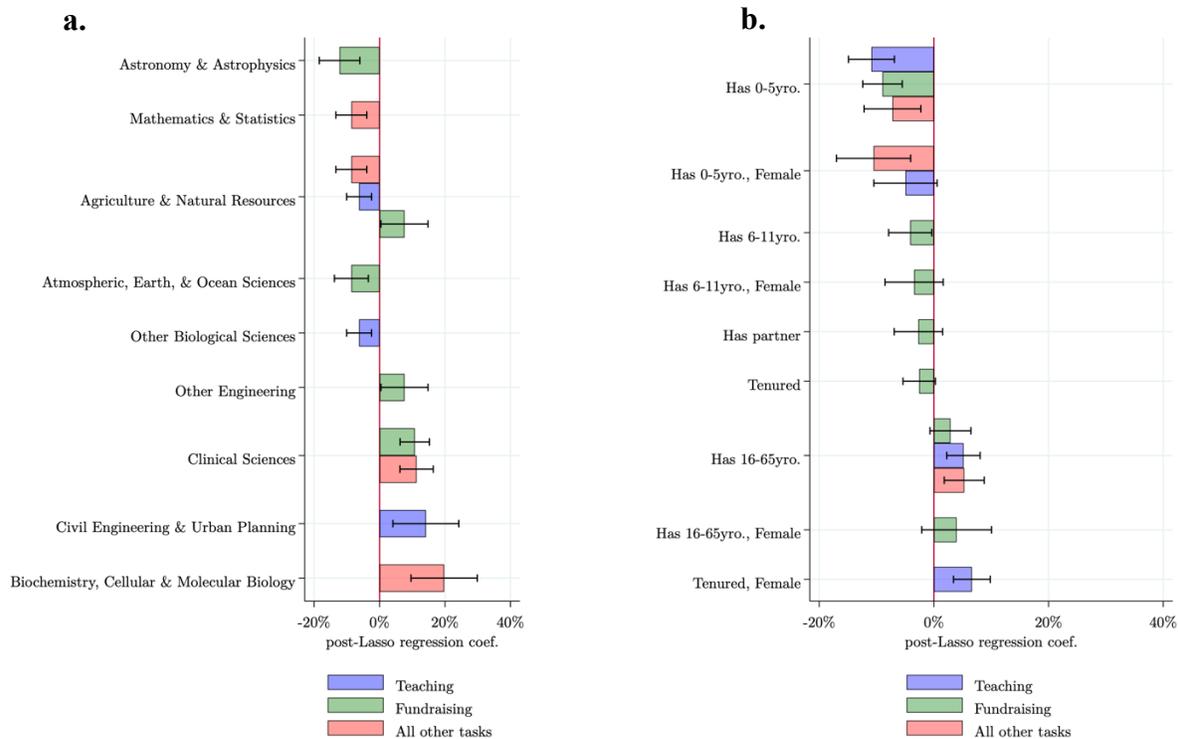

**Figure S5. Conditional changes in time allocations per Lasso selection. a.** Replicates Figure 2.b. from the main text, for each of the three non-research task categories. **b.** Replicates Figure 2.d. from the main text, for each of the three non-research task categories. In both panels, as in Figures 2.b. and 2.d, the error bars indicate 95% confidence intervals, and only variables selected in the corresponding Lasso selection exercises are included in the post-Lasso regression.

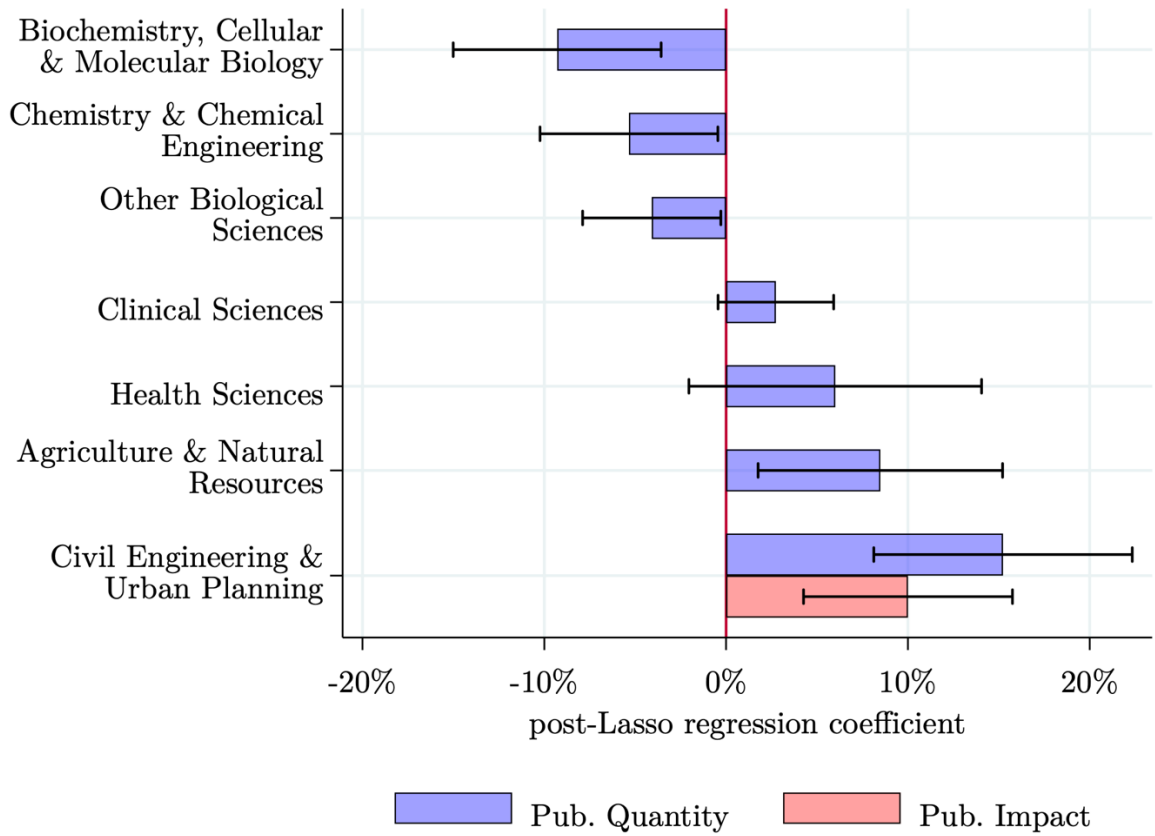

**Figure S6. Conditional changes in forecasts per Lasso selection, by field.** Replicates Figure 3b from the main text, instead focusing on fields of study. The error bars indicate 95% confidence intervals, and only variables selected in the corresponding Lasso selection exercises are included in the post-Lasso regression.

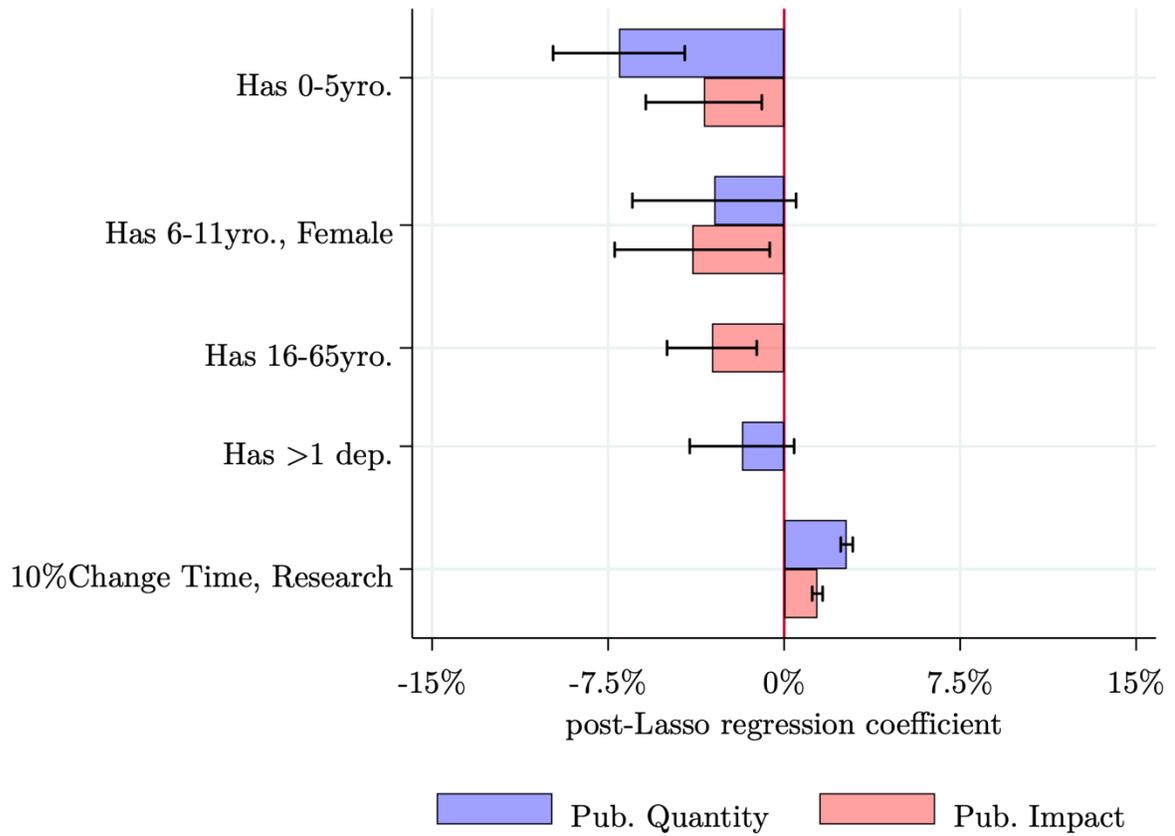

**Figure S7. Conditional changes in forecasts per Lasso selection, including time changes, by group.** Replicates Figure 3b from the main text, also including the scientists' reported changes in time committed to each of the four task categories. The error bars indicate 95% confidence intervals, and only variables selected in the corresponding Lasso selection exercises are included in the post-Lasso regression. The coefficient corresponding to the "10%Change Time, Research" variable indicates the percent change in the scientists' forecasted quantity or impact associated with a 10% increase in the change in reported research time. For example, we estimate that scientists who reported a 10% larger decline in their research time forecast that the pandemic will cause them to produce 2.67% fewer publications in 2020-2021.

# S8 References for Supplementary Information